\documentclass[12pt]{article}
\usepackage[english]{babel}
\usepackage[letterpaper,top=2cm,bottom=2cm,left=3cm,right=3cm,marginparwidth=1.75cm]{geometry}
\usepackage{graphicx}
\usepackage{multirow}
\usepackage{booktabs}
\usepackage{wrapfig}

\usepackage{titlesec}

\usepackage{amsthm}
\newtheorem{theorem}{Theorem}

\usepackage{titlesec}
\titlelabel{\thetitle.\quad}

\title{RDNF Oriented Analytics to Random Boolean Functions}         
\author{Levon Aslanyan, Irina Arsenyan, Vilik Karakhanyan, Hasmik Sahakyan}
\date{Institute for Informatics and Automation Problems \\ email: kavilik@gmail.com}          % Enter your date or \today between curly braces
\begin{document}

\maketitle
\begin{center}
{Abstract} 
\end{center}

Dominant areas of computer science and computation systems are intensively linked to the hypercube-related studies and interpretations. This article presents some transformations and analytics for some example algorithms and Boolean domain problems. Our focus is on the methodology of complexity evaluation and integration of several types of postulations concerning special hypercube structures. Our primary goal is to demonstrate the usual formulas and analytics in this area,  giving the necessary set of common formulas often used for complexity estimations and approximations. The basic example under considered is the Boolean minimization problem, in terms of the average complexity of the so-called reduced disjunctive normal form (also referred to as complete, prime irredundant, or Blake canonical form). In fact, combinatorial counterparts of the disjunctive normal form complexities are investigated in terms of sets of their maximal intervals. The results obtained compose the basis of logical separation classification algorithmic technology of pattern recognition. In fact, these considerations are not only general tools of minimization investigations of Boolean functions, but they also prove useful structures, models, and analytics for constraint logic programming, machine learning, decision policy optimization and other domains of computer science.

Keywords: Boolean function, hypercube, complexity, asymptotic, reduced disjunctive normal form.

\section{Hypercube and Related Structures}
\label{sec1}
The metric theory of Boolean functions (BF) \cite{Zhu62}, \cite{Yab74} arose in the 70's, in parallel with the emergence of broader design and implementation ideas for mechanical and electronic computation devices. It was then that it turned out that the system of binary representation of numbers is the most optimal, both from the point of view of the algorithmic implementation of arithmetic calculations and also from the point of view of developing physical carriers of performing these calculations \cite{Kit58}. BF -- functions with only binary variables, and also with values in the domain $\{0,1\},$ although simple among the other similar mathematical concepts, they are quite complex in solving problems associated with their transformations and optimization. The metric theory of Boolean functions provides the necessary knowledge for coding, transforming and implementing binary functions. Although the way to minimal BF representations are and remains difficult, a rather complete picture of the main forms of function representation of functions has been obtained, and the basic role here takes the concept of disjunctive normal forms. Successive steps of several transformations of functions are found to achieve minimal forms as a chain from the table or formula representation to the reduced d.n.f., then to the deadlock forms and finally -- the minimal structures. The accompanying structures and bottlenecks of achieving acceptable optimization are investigated intensively \cite{Zhu62}, \cite{Vas66}, \cite{Gla67}, \cite{Sap67}, \cite{Lup58}.
Here we will not cover the whole theory but will pay attention to one fundamental construction, -- to the concept of reduced disjunctive normal forms (r.d.n.f.) of Boolean functions. R.d.n.f. is the collection of all minimal conjunctions and geometrically - the system of all maximum intervals/sub-cubes of functions. These forms are a universal concept, and they also arise in problems such as circuit design from set of functional elements (logical part of chip design), in the theory of pattern recognition (logic separation algorithm, and generation of logical regularities) \cite{Asl75, Asl76, Asl07, Asl08}, in biological models of heredity and mutations (phylogeny, parsimony) \cite{Asl15, Asl17}, etc. Turning to the complexity characterization of structures associated with the reduced disjunctive normal form, where two types are usually considered: the largest and most typical characteristics, we will focus on the second component. In a concise survey of the domain, the initial studies of \cite{Gla67}, \cite{Mil64}, and \cite{Mil65}, should be mentioned, that give the formulas of average numbers of maximal intervals in Boolean functions. \cite{Asl74a}, \cite{Asl74b} extended these results to the case of partially defined Boolean functions. An alternative track of papers in these topics includes the articles \cite{Sko86}, \cite{Tom83}, \cite{Web83}. Current research on the topics of BF and complexities might be demonstrated through the papers \cite{Gar05},  \cite{Boy00}, \cite{Gon12}, \cite{Hru00},  \cite{Gui22}, \cite{Cha20}.
Methodologically, in studies in the area of BF, it should be taken into account that the function determination domain, as well as the number of functions itself, are finite, depending on the number of the variables -- the dimensionality. So, considering the parameter $\pi(f)$ over the functions, we get the split of these functions into finite classes by the values of this parameter. These are the rates and intensity of the accepted values of the parameter $\pi(f)$. In some cases, it is convenient to refer to these valuations as probabilistic distributions, which is not obligatorily but is convenient in some contexts. In this concern, there appears a link to the model of Random Boolean functions and the combinatorial theories initiated by A. Renyi and P. Erdos \cite{Erd59}, \cite{Erd63}.

\subsection{Concepts and definitions in the binary domain}
\label{ssec1.1}

\textbf{\textit{Elementary conjunction, Direction.}} Let $\tilde \alpha$ and $\tilde \beta$ -- be arbitrary vertices of the $n$-dimensional unite cube. And let $j_i, i = 1, 2, \cdots, r$ be all coordinates, those where $\alpha_{j_i} = \beta_{j_i}.$ Consider the formula 

$${\cal K}
(x_1, x_2, \cdots, x_n) = \bigwedge_{i=1}^r x_{j_i}^{\sigma_{j_i}},
$$ 

with $\sigma_{j_i} = \alpha_{j_i},$ $i = 1, 2, \cdots, r.$ We say that $K$ is an elementary conjunction stretched on the pair of vertices $\tilde \alpha$ and $\tilde \beta$ of the $n$-dimensional unit cube $E_n.$ The number of literals in $\cal{K}$ is the rank of $\cal{K}.$ The geometrical counterpart of $\cal{K}$ is a sub-cube defined as follows. Assign $0$ values to all but $j_1, j_2, \cdots, j_r$ coordinates and denote this vertex by $v_0.$ Similarly, assign these coordinates by the value $1$, obtaining the vertex $v_1.$ These are the minimal and maximal vertices that belong to $\cal K$,  and they determine a unique sub-cube of all truth vertices of $\cal K$. $n-r$, the number of variable coordinates of $\cal K$ is the size of its sub-cube.

Let $\lambda = \{j_1, j_2, \cdots, j_r\}$ be a collection of $r$ indices drawn up of variables $x_1, x_2, \cdots, x_n,$ and let $\bar\lambda$ be the complementary to the $\lambda$ set of indices. Conjunctions of the form $\bigwedge_{i=1}^r x_{j_i}^{\sigma_{j_i}}$ and the corresponding intervals will be called conjunctions and intervals of the direction $\lambda.$  For a fixed $r$ there are $C_n^r$ different directions, and each of them is determined by the appropriate  selection of an $r$ subset $\{j_1, j_2, \cdots, j_r\} $ of the set $\{1,2,...,n\}.$ The individual interval in the direction  $\{j_1, j_2, \cdots, j_r\}$ appears in result of assigning the values $\sigma_1, \sigma_2, \cdots, \sigma_r$ to the variables $x_{j_1}, x_{j_2}, \cdots, x_{j_r}.$ 

\begin{wrapfigure}{l}{0.7\textwidth}
  \begin{center}
    \includegraphics[width=0.48\textwidth]{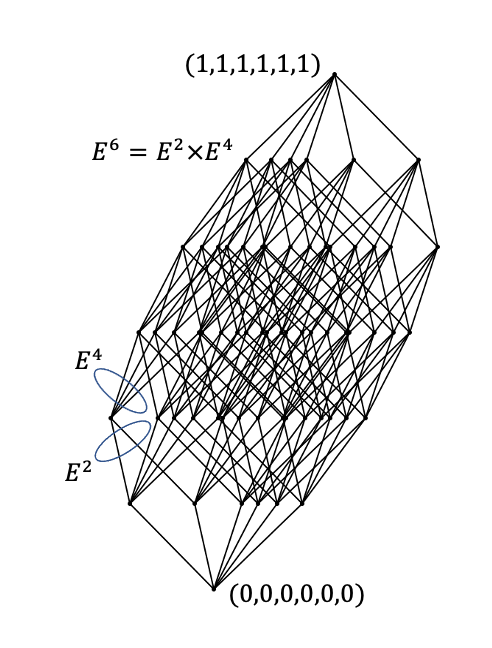}
  \end{center}
  \caption{Geometry of hypercube}
\end{wrapfigure}

\bigskip
This also means that there are $2^{n-r}$ conjunctions and intervals in one of the $r$-directions. The collection $\bar\lambda$ of indices defines another set of directions.

\bigskip
Let $\cal F$ be an arbitrary logical formula and ${\cal M} \subseteq B^n.$ We say that $\cal F$ absorbs or covers $\cal M$ if on each tuple $\tilde \alpha \in {\cal M}$ the formula $\cal F$ accepts the unite (true) value. \par
\bigskip
Let $\tilde \alpha \in E^n$ be an arbitrary vertex. Call the value $\mid \tilde \alpha \mid = \sum_{i=1}^n \alpha_i$ the module or the weight of $\tilde \alpha.$ The set of all vertices $\tilde \beta \in E^n,$ with $\rho(\tilde \alpha, \tilde \beta) =\mid~\tilde \alpha~\oplus~\tilde \beta~\mid = k,$ call the $k$--th layer of $E^n$ in relation to the vertex $\tilde \alpha\ $ ($\oplus$ -- mentions  $mod2$ summation).
\par
\bigskip

Intervals $\cal N_{K^1}$ and $\cal N_{K^2},$

$$K^1(x_1, x_2, \cdots, x_n) = \bigwedge_{i=1}^r x_{j_i}^{\sigma_{j_i}^1}  \ and \ K^2(x_1, x_2, \cdots, x_n) = \bigwedge_{i=1}^r x_{j_i}^{\sigma_{j_i}^2}$$ 

of the same size and the same direction we call neighbors if $\rho(\tilde \sigma^1, \tilde \sigma^2) = 1,$ where $\rho$ -- be the Hamming distance, $\rho(\tilde \sigma^1, \tilde \sigma^2) = {\sum_{i=1}^r \mid\sigma_{j_i}^1 - \sigma_{j_i}^2~\mid.}$ Let then $j_{i_0}$ is the number of that unique coordinate for which $\sigma_{j_{i_0}}^1 \ne \sigma_{j_{i_0}}^2.$ Then we say that the conjunctions $K^1$ and $K^2$ (or the pair of neighbor intervals corresponding to them) joined by the coordinate $x_{j_{i_0}},$ and, as a result, a new conjunction (interval) appears:
$$
\bigwedge_{i \ne i_0, i=1}^r x_{j_i}^{\sigma_{j_i}}.
$$

Partition the variable set $x_1, x_2, \cdots, x_n$ in an arbitrary manner into two nonempty groups:
$x_{i_1}, x_{i_2}, \cdots, x_k$ as the first group, and $x_{i_{k+1}}, x_{i_{k+2}}, \cdots, x_{i_n}$ as the second. 
Then, the $n$-dimensional unit cube $E_n$ may be represented as the Cartesian multiplication $B^k \times B^{n-k}$ of two sub-cubes: $B^k$ and $B^{n-k}$ generated correspondingly by the sets of variables $x_{i_1}, x_{i_2}, \cdots, x_{i_k}$  and $x_{i_{k+1}}, x_{i_{k+2}}, \cdots, x_{i_n}$. Let us enumerate the vertices of $B^{n-k}$ by the layers relative to the vertex $\tilde 0$ of $B^{n-k}$. Enumeration among the vertices of a particular layer is arbitrary, but the first group that is enumerated by low numbers is layer zero, then the first layer, and so on. Additional ordering among layer vertices may use lexicographic order, binary value based order, etc.

Consider an arbitrary $k$-dimensional sub-cube $B^k$ of $E^n,$ the first  $k$-dimensional interval $B^k_1$ in the direction of $B^k$. List the neighbor intervals to the considered one, $B^k_1,$ - $B^k_2, B^k_3, \cdots, B^k_{n-k+1}.$ Let $f$ be an arbitrary (partially defined) Boolean function that satisfies the following conditions:

\begin{itemize}

\item[$\alpha)$]
\ $B^k_1$ doesn't contain zero value vertices of $f:$ \ ($\forall \tilde \alpha \in
B^k_1, \ f(\tilde \alpha) \ne 0$),

\item[$\beta)$]
\ Each of the neighbor with $B^k_1$ interval contains at least one  `unit' value vertex $f:$ \ $(\forall j,\ j=2, 3, \cdots, n-k+1 \ \exists
\tilde \alpha \in B^k_j, \ f(\tilde \alpha) = 1),$

\item[$\gamma)$]
		\ $B^k_1$ \ contains at least one `unit' vertex of $f:$ \ $(\exists \tilde \alpha \in B^k_1, f(\tilde \alpha) = 1).$ \end{itemize}

\noindent
In conditions $\alpha), \ \beta), \ \gamma)$, we say that $B^k_1$ is a maximal interval of the function $f.\ $ d.n.f., composed of all elementary conjunctions, corresponding to maximal intervals of function $f$ is named the reduced disjunctive normal form of $f.$ The number of disjunctive members of this formula is considered as its complexity. Denoting by $r_k(f)$ the number of all maximal $k$--intervals of the function $f$ we get the formula of complexity of the reduced disjunctive normal form of $f$: 
$$
\sum_{k=0}^n r_k(f).
$$

\section{On the Maximum Number of $k$-Dimensional Maximal Intervals of RBF}
\label{sec2}

Consider the class $\ P_2(n)\ $ of all Boolean functions of $\ n\ $ variables $ x_1, x_2, \cdots, x_n$.  Let $\ p,\ 0<p<1\ $ be a fixed number, and $\ F_p\ $ -- the probability distribution on $\ P_2(n),\ $ generated in the following way. The function $\ f \in P_2(n)\ $ is induced as a result of a randomized experiment, where the values of the function on vertices of $\ E^n$ are derived randomly. The value $\ 1\ $ appears with a probability $\ p\ $ and the $\ 0\ $ value -- with a complementary probability $\ 1-p.\ $ The vertices of $\ E^n\ $ take part in this experiment independently of each  other, and the probabilistic distribution $\ F_p\ $ over the set of Boolean functions is generated in this way.
The probability of an individual Boolean function $f$ under the distribution $\ F_p\ $ depends on the balance between the $0$ and $1$ values of the function $f$ (the volumes of the sets $\cal N_f$ and $E^n - \cal N_f$). For $f \in P_2(n)$, this probability is equal to $p^{|\cal N_f|}(1-p)^{2^n-{|\cal N_f|}}$. When $p=1/2$ this probability is simply $1/{2^{2^n}}$and the corresponding distribution becomes the uniform distribution over the $\ P_2(n)$.
We introduce the notation$\ r_k(f) \ $ for the number of $\ k$-dimensional maximal intervals of the function $\ f \in P_2(n). $ And let $r_k(n,p)$ be the average value of the number of $\ k$-dimensional maximal intervals of functions $\ f \in P_2(n) \ $ under the distribution $\ F_p. \ $ It is easy to make sure, that
\begin{equation}
r_k(n,p) = \sum_{f \in P_2(n)} F_p(f)*r_k(f) 
\label{eq:1}
\end{equation}
The number $r_k(n,p) \ $ in the expression (\ref{eq:1})  is given by its definition as a sum over all functions of $f \in P_2(n), $ counting all their $k$-dimensional maximal intervals and taking into account the probabilities of $f$ in the distribution $F_p.$

\begin{figure}
Further evidence of these constructions is provided by the following scheme:
\centering
\includegraphics[width=1.0\textwidth]{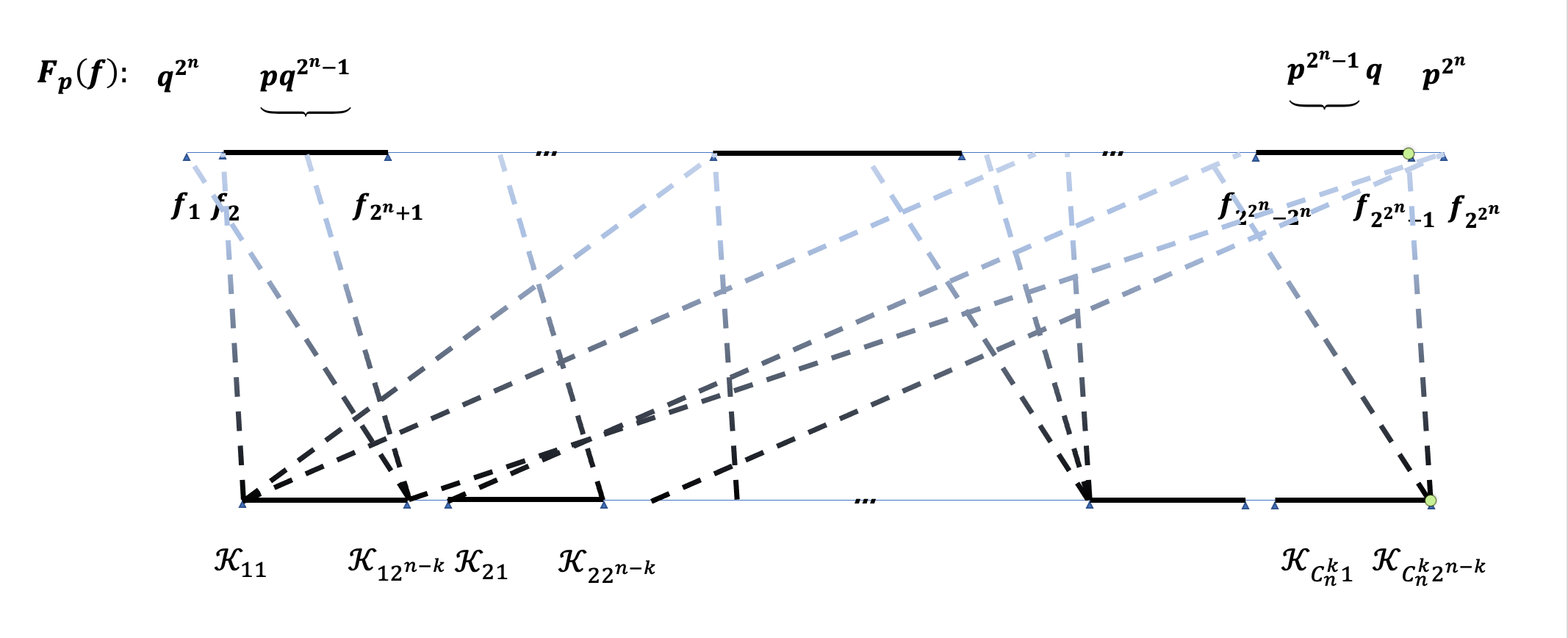}
\caption{\label{fig:maximality}{This figure presents the bipartite graph of functions and $k$-dimensional maximal intervals. Upper line functions are placed in order of the number of their "true" values, from $0$ to $2^k.$} Different functions include different numbers of $k$-dimensional maximal intervals and have different probabilities under the distribution  $\ F_p. \ $Instead, each interval presented in the bottom line is connected to the same number of functions. This is because  the sizes of intervals is the same. The order of intervals is by groups of intervals, that belong to the same direction. Numeration inside the functions with the same number of "ones" and inside the groups of intervals of the same direction is arbitrary.}
\end{figure}

Following \cite{Gla67}, we change the order of counting in \ref{eq:1}, first considering all $k$-dimensional intervals in $E^n$. We relay two events to these intervals: the one, about their maximality, and then the second, about the set of functions that accept the first event about maximality. In this regard, it is also convenient to split the$\ E^n$ in parts: the current $k$-dimensional interval $\cal K$ and its all $n-k$ neighboring $k$-dimensional intervals ${\cal K}_1, {\cal K}_2, \cdots, {\cal K}_{n-k}.$ This part, the current interval and its neighbors, covers an area ${\cal E}_1$ of ${2^k}(n-k+1)$ vertices of $\ E^n.$ And the second part that we consider, consists of the complementary area ${\cal E}_2$ to ${\cal E}_1$ up to $\ E^n.$ The probability of maximality of $\cal K$ for the function $f$ becomes the product of probability of maximality of $\cal K$ together with the conditional probability of $f$ when $\cal K$ is given to be maximal. The first probability equals $p^{2^k}(1-p^{2^k})^{n-k}.$ The first and second parts consist of events, and their sums of probabilities are equal to 1 as a probabilistic distribution. Now, when we sum up the mentioned conditional probabilities with all $f,$ we get the probability 1, and the final probability of maximality of $\cal K,$ under the conditions of $F_p,$ becomes $p^{2^k}(1-p^{2^k})^{n-k}.$ It reminds us to take this probability for all $k$-dimensional intervals, obtaining the following equivalent form for (\ref{eq:1}), 

\begin{equation}
r_k(n,p) = C_n^k 2^{n-k} p^{2^k}{(1-p^{2^k})}^{n-k}.
\label{eq:2}
\end{equation}
\begin{theorem} 
$r_k(n,p)$ is a concave function of the parameter $k$ in the interval $[0,n]$.
\end{theorem}

It is important to know the behavior of the function $r_k(n,p$ defined on the interval $[1,n]$. Initially, it is useful to calculate the values of the function at the boundary points of the domain of definition: $k=0,1,...,n-1,n$. We give these values both for the arbitrary $p$ and the value $1/2$.

 \begin{figure}
 \begin{center}
 \begin{tabular}{ |p{3cm}||p{4cm}|p{5cm} |}
 \hline
 \multicolumn{3}{|c|}{Boundary point values of {$r_k(n,p)$}} \\
 \hline
 Dimension $k$ of maximal interval & $r_k(n,p)$ & $r_k(n,1/2)$\\
 \hline
 $k=0$   & $2^n p(1-p)^n$  & $1/2$ \\
 $k=1$   & $n2^{n-1}p^2(1-p^2)^{n-1}$  & ${(n/4)}{{(3/2)}^{n-1}}$ \\
  $...$   & $...$  & $...$ \\
 $k=n-1$   & $n2^{n-1}p^{2^{n-1}}(1-p^{2^{n-1}})$  & $n2^{n-1}(1-{1/2^{2^{n-1}}})/2^{2^{n-1}})$ \\
 $k=n$   & $p^{2^n}$  & $1/{2^{2^n}}$ \\
 \hline
 \end{tabular} \\
 \caption{\label{fig:boundaypoints}}{Values of $r_k(n,p)$ on boundary points, such as $k=0,1,...,n-1,n.$}
 \end{center}
 \end{figure}

 As we can see, both the left and right boundary point values of the interval $(0,n)$ are small, but there is a noticeable increase from left to right at the left end, and a decrease from left to right at the right end. To get a complete picture of the behavior, consider a number of special intermediate point values of the function at:
 $$\ k_1 = log{\frac{1}{-logp}}, \  k_0=log{\frac{logn}{-logp}}, \ and \ k_1=log\frac{n}{-logp}.$$
 The technical element of choosing of these values is in simple evaluation of sub-formula $E_k=2^{2^k}$, which is an important functional part of the \ref{eq:1}. Substituting   $k_1, \ $ $k_0,$ and $k_2 \ $ into $E_k$ we get:
 \begin{equation} 
 E_{k_1}=1/2, \ E_{k_0}=1/n, \ E_{k_2}=1/2^n.
 \label{eq:3}
 \end{equation}
Let us start the proof of postulations 1-3. For this, conduct a preliminary analysis of the expression (\ref{eq:2}) for $\ r_k(n,p). \ $ Consider an arbitrary integer value function $\ k(n)$ that obeys the restriction $0 \leq {k(n)} \leq n,\ $ and  substitute it into the expression \ref{eq:2}. We are interested in the behaviour of the received function $\ r_{k(n)}(n,p) \ $ depending on the parameter $\ k(n) \ $ as $\ n\to\infty.$ \par

First let's make sure that with the increase of $k \ $ the expression $r_k(n,p) \ $ increases monotonically by the $\ k \leq [k_0], \ $ and then it decreases, when $\ ]k_0[ \leq k.$
By doing this we compose the relation
\begin{equation}
R_k=\frac{r_{k+1}(n,p)}{r_k(n,p)} =
\frac{(n-k)p^{2^k}(1+p^{2^k})^{n-k}} {2(k+1)(1-p^{2^{k+1}})}.
\label{eq:4}
\end{equation}
This expression can be considered for an arbitrary (not only for the integer) assignment to the parameter $k. \ $We will follow by checking if this function is concave in the interval $0<k<n$ for large $n.$ The direct way of this is to derive the expression of the fraction $R_k$ and treat it for a possible constant/zero value of it. In such consideration, the most important role takes the part $A_k=(n-k)p^{2^k}$ of the base expression \ref{eq:4}. Substituting $k_0$ into $A_k$ we obtain that $(n-k_0)p^{2^{k_0}}=(n-k_0)p^{(\frac{logn}{-logp})}=(n-k_0)2^{logp(\frac{-logn}{logp})}=\frac{n-k_0}{n}$, which is converging to 1 as  $n\to\infty.$ 
With the help of formulas in Section~\ref{sec3} we see that the part $B_k=(1+p^{2^k})^{n-k}$ of (\ref{eq:4}) is limited at the point $k_0$: (\ref{eq:3.1})  gives $(1+p^{2^{k_0}})^{n} \to e$ as $n \to \infty,$ so that $(1+p^{2^{k_0}})^{n-k_0}$ also tends to $e$. 
Compose the fraction $B_{k+1}/B_k$ in the following form:

\begin{equation}
B_{k+1}/B_k=\frac{\left({1+{{p^{2^k}}{p^{2^k}}}} \right)^{n-k-1}} {\left({1+{{p^{2^k}}}} \right)^{n-k}}=
\frac{\left(\frac{1+{{p^{2^k}}{p^{2^k}}}} {{1+p^{2^k}}}\right)^{n-k}} {1+{p^{2^k}}{p^{2^k}}}
\label{eq:5}
\end{equation}

\begin{wrapfigure}{l}{0.6\textwidth}
  \begin{center}
    \includegraphics[width=0.48\textwidth]{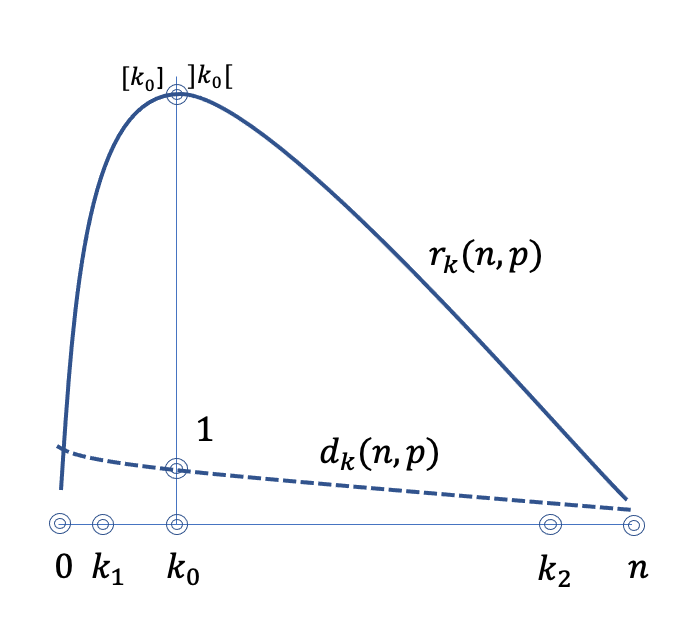}
  \end{center}
  \caption{Differential of growing $r_k(n,p).$}
\end{wrapfigure}

Note that the fraction $\frac{1+{{p^{2^k}}{p^{2^k}}}} {{1+p^{2^k}}}$ is less than $1,$ so its $n-k$ degree is also less than $1.$ And the denominator of (\ref{eq:5}) is greater than $1$ so that, finally, the expression (\ref{eq:5}) is less than $1$ for all $k$, which means a monotonic decrease of the expression $R_k$ in (\ref{eq:5}).
In general, as $\ k \ $ increases, all the factors of (\ref{eq:4}), other than $B_k,$ decrease monotonically and, besides this, as $\ n\to\infty \ $, this expression tends to zero at the point $\ k_0 \ $ and grows infinitely when $\ k=k_0-1.$ Finally, we receive that with increasing $\ k ,\ $for the beginning, $\ i_k(n,p) \ $ increases, achieving its maximal value at the point $\ [k_0]\ $ or $\ ]k_0[,\ $ and, then, it decreases.
\\
\section{On the Dependency of Number of $k$-Dimensional Maximal Intervals on $k$}
\label{sec3}

Consider the parameter$\ k_2=log{\frac{n}{-log{p}}}. \ $ Since $\ 0<p<1, \ $ we have $\ k_2=log{n+c}, \ $ where $\ c \ $ represents an absolute constant determined by the fixed value of $\ p. \ $ We intend to obtain an asymptotic formula for $\ i_k(n,p) \ $ by the $\ n\to\infty \ $for the values of $k$ of the form $\ k_2+const. \ $ We make use of the following expressions $\ C_n^k \sim \frac{n^k}{k!}, \ $ $\ (1-p^{2^k}) \sim 1,$ and $\ n! \sim n^n e^{-n} \sqrt{2 \pi n} \ $ as $\ n\to\infty, \ $ which are based on the formulas \\

\begin{itemize}
\item[$1.$]
\noindent
If \ $0 \le x \le 1$ \ and $0 \le y,$ \ then 
\begin{equation}
exp(x(1-\frac{x}{2})y) \le (1+x)^y \le exp(xy).
\label{eq:3.1}
\end{equation}

\item[$2.$]
\noindent
If \ $0 \le x \le 1$ \ and $0 \le y,$ \ then
\begin{equation}
(1-x)^y \le exp(-xy); \mbox{ and}
\label{eq:3.2}
\end{equation}
$exp(-x(1-x)y) \le (1-x)^y,$
when additionally $\ 0 \le x \le 1/2.$

\item[$3.$]
\noindent
If \ $x$ \ and \ $y$ \ be natural numbers, and $x \le y,$ then 
\begin{equation}
(1-\frac{x}{y})^{\frac{x-1}{2}} \le \prod_{i=1}^{x-1}(1-\frac{i}{y}) \le
(1-\frac{x}{2y})^{x-1}.
\label{eq:3.3}
\end{equation}
\end{itemize}
\par
and are valid for the mentioned values of the parameter $\ k, \ $ and for this reason 

\begin{equation}
i_k(n,p) \sim
\frac{n^k e^k 2^{n-k} p^{2^k}}{k^k \sqrt{2 \pi k}} = \tilde i_k(n,p).
\label{eq:3.4}
\end{equation}
\begin{theorem}
The probability, that functions of the class $\ P_2(n) \ $ under the distribution $\ F_p \ $ have maximal intervals of sizes $\ k,\ k<[k_1] \ $ or $\ k>[k_2], \ $ where $\ k_1 = log{\frac{1}{-logp}} \ $ and $\ k_2=log{\frac{n}{-logp}} \ $ tends to zero with $\ n \to \infty.$
\end{theorem}

On the right side of (\ref{eq:3.4})  we have expression, that depends on the continuous argument $\ k,$  and which is equivalent to the expression $\ i_k(n,p) \ $ for the integer values of the parameter $\ k, \ $ of the form $\ k_2+const. \ $ In the mentioned area, $\ \tilde i_k(n,p) \ $ decreases monotonically with the increase of $\ k,$ $\tilde i_{k_2}(n,p) \ $ tends to infinity, and $\ \tilde i_{k_2+1}(n,p)$ tends to zero, when $\ n\to\infty, \ $ so that $\ i_k(n,p) \to 0, \ $ for values $\ k>]k_2[ \ $ and $\ i_k(n,p) \to \infty \ $ for values $\ k_0 \leq k \leq [k_2], \ $ by $\ n\to\infty. \ $ Let us also denote, that  we do not insist that$\ i_{]k_2[}(n,p) \ $ as$\ n\to\infty $ converges to any appropriate value. \par
In what follows, we will use the first Chebyshev inequality (\ref{eq:1}). The first inequality lets formulate an extension of a postulation from \cite{Asl74c} for the case of the probability distribution $\ F_p. \ $ Actually, if to consider the expression $\ i_k(f), \ $ as a parameter of $\ \pi(f)\ $ then for the values $\ k>]k_2[ \ $ $\ i_k(n,p) \to 0 \ $ by $\ n\to\infty, \ $ and taking into the force the first inequality for the arbitrary $\ \epsilon(n) \geq 0 \ $ $\ P(i_k(f) \geq \epsilon(n)) \to 0 \ $ when $\ n\to\infty.$ \\
A similar situation takes place in the region of small values of the parameter $k$. For the value $k=k_1$ and $p=1/2$ by the (\ref{eq:3}) $p^{2^{k_1}}=1/2$ and $r_{k_1}(n,p) \to \infty$ as $n\to\infty.$ For $p>1/2$, already for the value $k_1-1$, we observe that  $r_{k_1-1}(n,p) \to 0$ as $n\to\infty.$ This is just because $\frac{2^{n-{k_1+1}}}{1-p^{2^{k_1-1}}}$is a decreasing exponent, which together with $C_n^k$ tends to $0$.

\section{Conclusion}
\label{sec4}

This article has two goals: first, it considers the set of formulas needed to analyze the complexity of structures associated with a multidimensional unit cube, providing the necessary transformations and approximations for these formulas. Further, the paper considers a typical study for this field using these formulas. The problem under consideration estimates the complexity of the reduced disjunctive normal form of Boolean functions on average, or, what is the same, for almost the entire class of functions.

\end{document}